# Investigation of methane adsorption and its effect on gas transport in shale matrix through microscale and mesoscale simulations

ZhongZhen Li [a], Ting Min [b,c], Li Chen [a,d], Qinjun Kang [d], Ya-Ling He [a], Wen-Quan Tao [a,*]

a: Key Laboratory of Thermo-Fluid Science and Engineering of MOE, School of Energy and Power Engineering, Xi'an Jiaotong University, Xi'an, Shaanxi, 710049, China b: State Key Laboratory for Mechanical Behavior of Materials, Xi'an Jiaotong University, Xi'an 710049, PR China

c: Materials Physics & Applications Division, Los Alamos National Laboratory, Los Alamos, New Mexico, USA

d: Earth and Environmental Sciences Division, Los Alamos National Laboratory, Los Alamos, New Mexico, 87545, USA

* Wen-Quan Tao, Email: wqtao@mail.xjtu.edu.cn

**Abstract**

Methane adsorption and its effect on fluid flow in shale matrix are investigated through multi-scale simulation scheme by using molecular dynamics (MD) and lattice Boltzmann (LB) methods. Equilibrium MD simulations are conducted to study methane adsorption on the organic and inorganic walls of nanopores in shale matrix with different pore sizes and pressures. Density and pressure distributions within the adsorbed layer and the free gas region are discussed. The illumination of the MD results on larger scale LB simulations is presented. Pressure-dependent thickness of adsorbed layer should be adopted and the transport of adsorbed layer should be properly considered in LB simulations. LB simulations, which are based on a generalized Navier-Stokes equation for flow through low-permeability porous media with slippage, are conducted by taking into consideration the effects of adsorbed layer. It is found that competitive effects of slippage and adsorbed layer exist on the permeability of shale matrix, leading to different changing trends of the apparent permeability.

## 1. Introduction

Shale gas refers to natural gas trapped within fine grained sedimentary rocks called mudstone or shale that are rich of oil/gas. Shale gas reservoirs have become major sources of natural gas production in North America, and are expected to play an increasingly important role in Europe and Asia in the near future. Over the past decades, advanced techniques such as horizontal drilling and multi-stage hydraulic fracturing have greatly promoted the exploitation of shale gas from shale matrix with low permeability [1]. With hydraulic fracturing, it is now generally agreed that the fabric of shale systems comprises primarily organic matter, inorganic matter, natural fractures and hydraulic fractures [2-10]. Such structure characteristics lead to multiscale pore systems in the shale gas reservoir and different transport mechanisms at different scales including viscous flow, slip flow, Knudsen diffusion and adsorption and desorption [2].

Recently, a great effort has been devoted to experimentally understand the geometrical characteristics of shale matrix. Particularly, advanced direct measurement techniques, such as scanning electron microscopy (SEM) combined with focused ion beam (FIB) milling, allow visualization of the nanoscale structures of shales on high quality flat surfaces, and provide new insights in the micro/nano-scale structures of shale matrix [3-8, 11, 12]. Interparticle pores, intraparticle pores and organic matter pores have been observed in shale matrix, with diameter from a few nanometers to a few micrometers [8]. The organic matter, widely known as kerogen, usually presents as discrete grains randomly imbedded within the inorganic matrix. The weight percentage of the organic matter in shale matrix is called total organic content (TOC), which is an important indicator of shale gas reserve. Shale gas is stored in shale matrix as free gas in the void space as well as adsorbed gas on the solid surface of pores. It is found that the amount of both adsorbed gas and free gas increases as the TOC increases [4]. In some shales, most of the porosity can be contributed by nanosize pores in the organic matter [3, 7]. These nanosize pores have significant internal surface areas, resulting in substantial adsorption of high-density shale gas on the pore walls [13]. It was estimated that between 20~85 % of natural gas in shale gas reservoir can be stored in the form of adsorbed gas [14].

The effect of the adsorbed gas on the fluid flow can be neglected for conventional rocks with large pore size. However, it should be considered in shale matrix where nanosize pores are dominant [9, 15-25]. Adsorption will affect the fluid flow in a nanosize pore through two mechanisms [16]. On one hand, in a nanosize pore gas slippage on the solid surface occurs. The adsorbed layer, sandwiched between the walls and the free gas region in the middle of the pores, will change the interactions between gas molecules and the wall, and thus greatly affects the slippage [16, 18]. On the other hand, the adsorbed methane occupies a significant part of the void space of a nanopore [9]. For example, 29.4% of the void space of a cylinder with a radius of 5 nm is occupied by the adsorbed layer with a thickness of 0.7 nm based on the molecular dynamics simulation results in Ref. [9].

In this work, multiscale simulations are performed to investigate methane adsorption and its effects on fluid flow in shale matrix by using Molecular dynamics (MD) method and lattice Boltzmann (LB) method. MD simulations are conducted to investigate the methane adsorption in nanopores of shale matrix with different solid surfaces and under different pressures, the results of which are then upscaled into larger scale LB studies. The remaining part of the present study is as follows. In Section 2, equilibrium MD simulations are performed to study the adsorption phenomenon of methane on the walls of the nanosize pores in shale matrix. Effects of the pore size, pressure and solid wall characteristics are explored in detail. In Section 3, the illumination of the MD simulation results on the investigation at a larger scale is discussed, and upscaling the MD results to the LB simulations is presented. Consequently in Section 4, a generalized LB model for fluid flow through tight porous media with slippage, which is developed in our previous work [19], is further improved to incorporate the effects of the adsorbed layer. The improved LB model is then employed to investigate the competitive effects of gas slippage and adsorption layer on permeability of shale matrix. Finally, some conclusions are drawn in Section 5.

## 2. Molecular dynamics simulation of methane adsorption

*2.1 Brief introduction of Molecular dynamics simulation*

MD simulations provide a methodology for detailed microscopic modeling on the molecular scale [26]. In a MD simulation, each atom or molecule of the matter studied is treated as a point mass and Newton's second law equation is integrated to compute their motion

$$\vec{F} = m\vec{a} = m\frac{d^2\vec{r}}{dt^2} \tag{1}$$

where $m$ is the mass of the site, $\vec{a}$ is the acceleration, $\vec{r}$ is the position vector and $t$ is the time. The force $\vec{F}$ is derived from the potential energy

$$\vec{F} = -\nabla E \tag{2}$$

The total potential energy $E$ contains two parts: intermolecular term and intramolecular term. The intramolecular term, $E_{bond}$, represents the interactions between the bonded atoms within the same molecule including the bond stretch, angle bend and torsion energy terms; whereas the intermolecular term, $E_{nonbond}$, is used to describe the interactions between atoms from different molecules that have contributions from the electrostatic (Coulombic) and the short-range (van der Waals) interactions. Therefore, the expression of potential energy $E$ is as follows

$$E = E_{bond} + E_{nonbond} = E_{\text{bond stretch}} + E_{\text{angle bend}} + E_{\text{torsion}} + E_{\text{Coul}} + E_{\text{VDW}} \tag{3}$$

A variety of useful microscopic and macroscopic information such as transport coefficients, thermodynamic properties and structural or conformational properties can be extracted from the trajectory of atoms' motion [27].

*2.2 Simulation setup and procedure*

Loucks et al. [8] classified pores in shale matrix into three types: interparticle pores that are found between particles and crystals of inorganic minerals, intraparticle pores within the particles, and organic-matter pores that are located within the organic matters. In the present study, MD simulations are performed to simulate methane distributions in these pores under equilibrium conditions, and to investigate methane adsorption on the solid surface of organic matter and inorganic matter. Following the work of Ambrose et al. [9], the organic matter is assumed to be represented by graphite. Inorganic minerals in shale matrix are composed of predominantly clay, calcite, quartz, and pyrite [8, 12]. In the present study, calcite is studied, which is one of the most common carbonate minerals and the most stable polymorph of the calcium carbonate. The computational domain is a three-dimensional orthorhombic pore consisting of upper and lower walls made of graphite or calcite, as shown in Fig.1 and Fig.2, respectively.

Fig.1 shows the simulation model of methane molecules in a graphite pore. The pore width $H$ is defined as the distance between the inner most graphene planes. The size of the pores in the organic matter is in a wide range, from a few nanometers to hundreds of nanometers [3, 4, 6, 8]. In the present study, values of 1, 3.6 and 7.2 nm for $H$ are studied. In following sections, we will show that under a certain pressure, adsorption thickness will not change if pore size is greater than a critical value, and a 7.2 nm pore is sufficiently large for typical values of pressure during shale gas extraction. The *x* and *y* dimensions of the simulation box are 42.6×4.92 nm. In order to investigate methane adsorption in different pore pressures, a total of 1000 to 7000 methane molecules are employed during the simulations, and the number of methane molecules for a certain simulation case is determined based on the geometry size and the pressure in the slit-pore. For the interactions between molecules, only the van der Waals interactions are taken into account due to the weak electrostatic interaction, which are modeled using the Lennard-Jones (LJ) 12-6 form potential with a cut-off distance equal to 1.5 nm

$$E_{\text{VDW}}(r_{ij}) = \begin{cases} 4\varepsilon \left[ \left( \dfrac{\sigma}{r_{ij}} \right)^{12} - \left( \dfrac{\sigma}{r_{ij}} \right)^{6} \right] & r_{ij} < r_{cut} \\ 0 & r_{ij} \geq r_{cut} \end{cases} \qquad (4)$$

where $r_{ij}$ is the distance between atom *i* and *j*, and $r_{cut}$ denotes the cut-off distance. The energy and distance parameters of LJ potential are represented by $\varepsilon$ and $\sigma$, respectively. The methane molecule is modeled using a united-atom carbon-centered LJ potential (OPLS-UA force field). The energy and distance parameters used for fluid-fluid and solid-solid interactions are listed in Table 1 [28]. Besides, Lorentz-Berthelot mixing rules are used to determine the solid-fluid interactions.

Calcite crystal belongs to the hexagonal crystal system, R3c space group [29]. The lattice parameters are as follows: $a = b = 0.4988$ nm, $c = 1.7061$ nm, $\alpha = \beta = 90°$, and $\gamma = 120°$. The (1$\bar{1}$0) and (104) faces of calcite crystal shown in Figs. 2(a) and 2(b) are created by surface cleaving with the molecular modeling program Materials Studio 6.1 from Accelrys Software Inc. The dimensions of the simulation wall with (1$\bar{1}$0) and (104) faces are 19.96×5.12×1.37 nm$^3$ and 20.24×4.99×1.07 nm$^3$, respectively, as shown in Fig. 2(c) and Fig. 2(d). The potential model developed by Pavese et al.[30] is used for calcite and the OPLS-AA force field [31] is employed to model the methane molecules because the Coulombic interaction between calcite and organic is not negligible. The Coulombic energy is represented

$$E_{Coul} = \frac{q_i q_j}{4\pi\varepsilon_0 r_{ij}} \tag{5}$$

where $q_i$ and $q_j$ denote the partial charge of atom $i$ and $j$. $\varepsilon_0$ is the dielectric permittivity of vacuum. The interaction potential between calcite and methane is modeled by a force field designed for bio-inorganic interfaces [32]. The partial charges and LJ parameters are listed in Table 2.

The Large-scale Atomic/Molecular Massively Parallel Simulation (LAMMPS) [27] is used to perform the MD simulations. Since the crystal grows along with rigid sequence and orientation, atoms in calcite and graphite wall are set to be frozen completely during the MD simulation [33]. A 1.0 fs time step is chosen for a 2.0 ns simulation run which consists of equilibrium and production run. The fluid system confined between walls is maintained at a constant temperature $T$ of 353.15 K, with an NVT ensemble using Nose-Hoover thermostat.

After the equilibrium is reached, a production run is carried out for the density and pressure computations. In order to calculate the density and pressure profile across the pore, the pore is divided into a number of 0.04 nm bins and the properties are averaged in each bin. The density of methane in every bin is calculated as follows

$$\rho_{CH_4} = \frac{N_{bin} M_{CH_4}}{N_A V_{bin}} \tag{6}$$

where $N_{bin}$ is the number of methane molecules in a bin and $N_A$ is the Avogadro constant. $M_{CH4}$ denotes molecular weight of methane and $V_{bin}$ represents the volume of a bin.

In LAMMPS [27], the stress tensor for atom $i$ is given by the following formula, where $a$ and $b$ take on values $x$, $y$, $z$ to generate the 6 components of the symmetric tensor

$$S_{ab} = - \begin{bmatrix} mv_a v_b + \frac{1}{2}\sum_{n=1}^{N_p}(r_{1a}F_{1b} + r_{2a}F_{2b}) + \frac{1}{2}\sum_{n=1}^{N_b}(r_{1a}F_{1b} + r_{2a}F_{2b}) + \\ \frac{1}{3}\sum_{n=1}^{N_a}(r_{1a}F_{1b} + r_{2a}F_{2b} + r_{3a}F_{3b}) + \frac{1}{4}\sum_{n=1}^{N_d}(r_{1a}F_{1b} + r_{2a}F_{2b} + r_{3a}F_{3b} + r_{4a}F_{4b}) + \\ \frac{1}{4}\sum_{n=1}^{N_i}(r_{1a}F_{1b} + r_{2a}F_{2b} + r_{3a}F_{3b} + r_{4a}F_{4b}) + Kspace(r_{ia}, F_{ib}) + \sum_{n=1}^{N_p} r_{ia}F_{ib} \end{bmatrix} \tag{7}$$

In this equation, the first term on the right is a kinetic energy contribution from atom $I$; the second term is a pairwise energy contribution where $n$ loops over the $Np$ neighbors of atom $i$, $r_1$ and $r_2$ are the positions of the 2 atoms in the pairwise interaction, and $F_1$ and $F_2$ are the forces on the 2 atoms resulting from the pairwise interaction; and the third to eighth terms are contributions of similar forms for the $N_b$ bonds, $N_a$ angle, $N_d$ dihedral, $N_i$ improper, long-range Coulombic interactions and internal constraint forces, respectively. The pressure of a bin is calculated by

$$P_{bin} = -\frac{\sum_{n=1}^{N_{bin}}(S_{xx} + S_{yy} + S_{zz})}{3V_{bin}} \tag{8}$$

*2.3 Molecular Dynamics Simulation Results and discussion*

*2.3.1 General pressure and density distributions*

Fig. 3(a) and (b) shows the pressure and density profile in a 3.6 nm graphite pore with 1000 and 7000 methane molecules, respectively. The most important observation from the figure is the high density/pressure regions adjacent to the walls that are distinct from the flat density region in the middle, which is exactly due to the adsorption effects of the solid walls. A methane molecule near the wall is strongly controlled by the pairwise and electrostatic interactions from the wall. As a molecule is away from the wall, the effects of wall become increasingly weak and the interactions among molecules gradually enhance, and thus methane is more like in the bulk fluid region as shown by the flat distribution regions of the density and pressure in Fig. 3. From the figure, it can be found that at a certain distance from the wall, effects of the solid walls can be neglected (dash line for the case of 7000 methane molecules in Fig. 3). In the present study, the gas between a wall and the dashed line is called adsorption gas, while that between the dashed lines is called free gas. The averaged density and pressure of the free gas for a bunch of simulation cases are calculated, as listed in Table 3, and are compared with the experimental results of bulk methane from the NIST database, as shown in Fig. 4. It can be found that the simulated results are in good agreement with the experiments, which validates the MD simulations in the present study, indicating that the methane molecules away from the walls are located in the bulk fluid region and the free gas domain defined here is reasonable.

*2.3.2 Effect of pressure on methane adsorption*

Fig. 5 shows the density profile across the 3.6 nm graphite pore under different pressures. Considering the critical point of methane with pressure of 4.579 Mpa and temperature of -82.3 °C, most of the simulation cases are in supercritical conditions except the one with pressure of 2.6 Mpa. Under a lower pressure (2.6 Mpa), mono-adsorption layer is formed. As the pressure increases (above 10 Mpa, in which the methane is supercritical), the adsorption effect of the wall becomes stronger and a second adsorption layer is gradually formed. As shown in Fig. 5, the first molecular layer is located at the first 0.4 nm from the wall and the second molecular layer is located at the next 0.4 nm, in consistent with the molecular diameter of the methane of about 0.38 nm.

To more clearly understand the effects of the pressure on the methane adsorption, the average density in the first and the second layer is further calculated and compared with the density of the free gas, as shown in Fig. 6. All the densities increase as the pressure rises, as expected. The density ratio of the first layer to the free gas ranges from 6.5 to 1.3 for the pressure studied, while that for the second layer to the free gas is much lower about 1.5 to 1.07. Interestingly, the ratio between densities of the first/second adsorption layer and free gas decreases as the pressure increases. This is because as the pressure increases, more surface cites are occupied by the adsorbed molecules, and less free sites are available for adsorption. Therefore the adsorption process gradually slows down and approaches saturation.

*2.3.3 Effect of pore width*

The adsorption in narrow pores is studied. Fig. 7 shows the density distributions of 1000 to 3000 methane molecules in a 1.0 nm graphite pore. As illustrated in Fig. 3, the influence distance of the walls can be as high as 0.8 nm. Therefore, for the pore with diameter of 1 nm studied here, the methane in the entire pore will be affected by the wall interaction. This is confirmed in Fig. 7, where no free gas region is observed. Given the methane molecular diameter of 0.38 nm, there is no chance for the methane to form four density peaks as shown in Fig. 5 for the higher pressure case. Therefore, for higher pressures (large methane molecular numbers) in Fig. 7, a three peak density distribution is observed. In the literature, it is found that for a pore with width less than 2 nm, methane molecules are always affected by the molecule-wall interactions and behavior of the adsorbed molecules should be considered rather than the motion of free gas molecules [9], which is in consistent with the simulation results in Fig. 7.

The adsorption in wider pores is also studied, as shown in Fig. 8 where the pore width is 7.2 nm. For comparison, the density distribution of that in 3.6 nm pore under the approximately same pressure is also displayed in Fig. 8. It can be observed that for the same pressure, the density distribution in the 7.2 nm almost coincides with that in the 3.6 nm pore. In the literature [9], it was found that for pores with width of 1.96 and 3.6 nm, the methane density in the first layer is 404 kg m$^{-3}$ and 372 kg m$^{-3}$, respectively. This means that a nearly 50% reduction in pore size leads to only 8.6% density increase, indicating a moderate-level dependence of the density on the pore size. Our simulation results of the density profile for pores with widths of 3.6 nm and 7.2 nm further confirm that if the pore width is higher than a certain value under a certain pressure, its effects on the density distribution can be ignored, resulting in the same density in the adsorption layer as well as that in the free gas region.

*2.3.4 Effect of solid materials*

Fig. 8 also shows the density distribution in the calcite pores. It can be seen that the density in the adsorption layer for the calcite is much lower than that in the graphite pore. This is in consistent with the consensus that methane tends to adsorb on the organic matter surface than on the inorganic matter surface. The density near the wall for different surfaces of calcite crystal is also quite different, indicating the crystal surface has a great impact on the adsorption phenomenon.

**3. Upscaling**

MD simulation usually requires huge computational resources, and thus limits itself for practical applications with engineering interest. Therefore, upscaling the MD results into a larger scale study is highly necessary. Recently, there have several studies using the LB method, which is considered as a mesoscopic numerical method beyond the microscopic MD, for shale gas flow in the nano/micro-size pores of shale matrix [16-25] regarding the effects of gas slippage [19-21, 23, 24] and adsorbed layer [18, 22, 25]. The MD simulation in the present study helps us to understand the adsorption of methane on the pore walls in shale matrix. It also is of great importance to the development of the LB models. The illumination of our MD simulation results on the current LB simulation is discussed as follows based on the two ways the adsorbed layer affects the shale gas flow in pores of shale matrix：reducing the pore volume, and changing the interaction between gas and solid walls (thus altering the slip at the fluid-solid interface) [16].

*3. 1. The adsorbed layers*

In the LB simulations for studying effects of volume reduction due to adsorption, it is commonly assumed that methane adsorption follows the Langmuir isotherm, and thus a computational grid with width of a monolayer methane (about 0.38 nm) is simply subtracted from the pore space during the LB simulations [16, 25]. Obviously, this effect reduces the permeability of the shale matrix. From our MD simulations, it can be found that under low pressures one layer of adsorbed methane is presented; however, as the pressure increases, a second adsorption layer is gradually formed, as shown in Fig. 5, which is also found in the literature [9]. Given that the pressure in the shale gas reservoir is usually higher than 10 Mpa, the second layer adsorption is believed to take place, and this should be considered in the LB simulations. Besides, a pressure-dependent thickness of adsorption layers is more reasonable [15]. Further, the adsorption also is affected by the solid surface property, for example, the organic matter surface or the inorganic matter surface, which is also required to be considered in the LB simulations.

In addition, as the pore size is reduced to 1 nm, the fluid flow in the entire pore is affected by the adsorption effect, presenting three adsorbed layers, as shown in Fig.7. Current LB models still have difficulties to model flow with high *Kn* number, and therefore must be greatly improved to consider the high *Kn* number shale gas transport in pores with a few nanometers, where the wall adsorption effect dominates the entire pore.

*3.2. The interaction between adsorbed layer and walls*

Recent experimental results of Kang et al. [13], which demonstrated that the adsorbed layers of shale gas are moving along the solid surface, has been very helpful for developing the LB model for gas transport in nano/micro-pores of shale matrix taking into account the effects of adsorbed layer on slip flow at wall-gas interface. Based on the experiment, shale gas flow in a nano/micro-size pore is considered as a combined result of free gas flow inside the pore and surface transport of the adsorbed gas along the solid wall [13]. Therefore, Fathi and Akkutlu [18] developed a LB-Langmuir isotherm (LB-LS) model, in which the LB with slip boundary condition is adopted for free gas flow and LS is used to calculate the transport rate of the adsorbed gas at the solid surface. The pseudopotential model proposed by Shan and Chen [34, 35] is employed to consider the interactions between solid-gas and gas-gas, with which nonideal gas effects can be considered. Very recently, Ren et al. [25] proposed a different form of LB boundary slip condition based on the Langmuir slip model, and they demonstrated that their model can predict more reasonable physical behaviors compared to that in Ref. [18, 22]. All these LB simulations [18, 22, 25] demonstrated that surface diffusion is an important characteristic, and revealed that shale gas transport in nano/micro-pores is significantly affected by the interaction between adsorbed layer and wall as well as that between adsorbed layer and free gas.

In all these LB studies, it is stated that "*the gas–solid and gas–gas molecular interactions at the solid surface can be incorporated into the LB model through certain suitable boundary conditions*". Indeed, in the simulations the solid-fluid interaction has already been considered through the boundary conditions. As mentioned above, in Fathi and Akkutlu's work [18], the Shan-Chen pseudopotential model (herein called SC model) is employed to consider the fluid-fluid and solid-fluid interactions. With the fluid-fluid interaction included in the SC model, the nonideal characteristics of shale gas is considered [34, 35]. However, in their model, the solid-fluid interaction, which has already been incorporated through the slip boundary condition, was also considered in the SC model, leading to double-counting such interaction. A detailed discussion about this point is as follows.

The adsorbed layer greatly affects the interaction between solid wall and shale gas, thus altering the slippage at the solid-fluid interface. For simulating gas slippage flow using LB models, to the best of our knowledge, there are two schemes [34]. One scheme is to adopt the slip boundary condition to implicitly consider the solid-fluid interaction, which is the scheme that widely employed in LB models for slip flow in the literature [36]. The second scheme is employing the SC model, where the solid-fluid interaction is calculated, to predict the interfacial behaviors at the fluid-solid interface. The capability of the SC model of simulating slippage has been demonstrated in the literature [37, 38], and one can refer to a recent review for more details [34]. A denser or thinner fluid region at the solid-fluid interface, compared to the fluid flow in the middle of the pores, has been captured, which is believed to cause the slippage at fluid-solid interface [37]. In fact, the SC model within the LB framework and MD show strong similarity for considering fluid-fluid and fluid-solid interactions, with the former using a pseudoptential while the latter using the actual potential. Such similarity enables the SC model to simulate slippage. Thus, in the literature the SC model has been called a "supermolecular simulator", serving as an upscaling simulator of slippage compared to MD [34, 37].

As pointed out by Ren et al. [25], the higher velocity in the adsorbed layer than that of the free gas obtained by the LB studies in Ref. [18, 22] is unreasonable. Such unreasonable velocity filed predicted is largely due to the double count of the fluid-solid interactions. In the recent study of Ren et al. [25], the slip boundary condition is used and the solid-fluid interaction is not repeatedly counted, and thus their simulations predicted more rational physical behaviors, where the velocity of adsorbed layer region is lower than that of free gas region and the transition between the two regions is smooth. Recent MD simulations of shale gas slippage in the nanopores of shale matrix [23] also support the simulation results of Ren et al. [25]. We also performed nonequilibrium MD simulations for methane flow in graphite nanopores at different pressures for pore width $H$ = 5.0 nm, 10.0 nm and 15.0 nm. Fig. 9 shows the density and velocity profile across a 15.0 nm graphite pore, where solid line is the parabolic fitting result. The adsorbed layer with higher density near the solid walls can be clearly observed. Our MD simulation demonstrated that the adsorbed layer moves along the solid surface, as shown in the bottom image of Fig. 9, in consistence with the experiments [13], which can also be observed in Fig. 10 for velocities in pores with $H$=15.0 nm under different pressures and in Fig. 11 for pores with different widths. From Figs. 9-11, it can be seen that the velocity in the adsorbed layer is always lower than that in free gas region, which agrees with the LB and MD simulation results in the literature [23, 25].

**4 Lattice Boltzmann simulation**

Adsorbed layer, which lies between the solid wall and the free gas, affects the pore volume as well as alters the interactions between solid walls and shale gas as discussed above, and thus plays an important role on the gas flow in shale matrix. Based on discussions in Section 3, it can be found that studies on the influence of the adsorbed layer on the solid-gas interactions are rather scanty, and in the current few openly published studies argument exists about the relative transport of the adsorbed layer and the free gas. Thus such an influence still needs to be further studied to be upscaled to the LB simulations.

In the present study, we only consider the first effect of the adsorbed layer in our LB studies, namely the volume reduction of void space. Unlike previous studies where a constant adsorption thickness is considered, a pressure-dependent adsorption thickness is adopted in the present study based on the MD studies.

*4.1 The lattice Boltzmann model*

The generalized LB model proposed in our previous work [19], which is for fluid flow through low-permeability porous media with Klinkenberg's effect, is adopted to study the effects of adsorption layer. This model is briefly introduced as follows. Details of the generalized LB model can be found in Ref. [19]. The model combined the generalized Navier-Stokes equations proposed by Nihiarasu et al. [39] for isothermal incompressible fluid flow in porous media (Eq. (9)) and the second-order Beskok and Karniadakis-Civan's correlation to calculate the apparent permeability (liquid permeability) based on intrinsic permeability (liquid permeability) and Knudsen number

$$\nabla \cdot \mathbf{u} = 0 \tag{9a}$$

$$\frac{\partial \mathbf{u}}{\partial t} + (\mathbf{u} \cdot \nabla)\frac{\mathbf{u}}{\varepsilon} = -\frac{1}{\rho}\nabla(\varepsilon p) + \upsilon_e \nabla^2 \mathbf{u} - \frac{\varepsilon \upsilon}{k_a}\mathbf{u} + \varepsilon \mathbf{G} \tag{9b}$$

where $t$ is time, $\rho$ volume averaged fluid density, $p$ volume averaged pressure, $\mathbf{u}$ the superficial velocity, $\varepsilon$ the porosity and $\upsilon_e$ an effective viscosity equal to the shear viscosity of fluid $\upsilon$ times the viscosity ratio $J$ ($\upsilon_e = \upsilon J$). The third term on RHS of Eq. (9b) is the linear (Darcy) drag force. Note that nonlinear (Forchheimer) drag force is not considered because usually flow rate is extremely low in low-permeability tight porous media such as shale matrix [2]. Gas slippage is a phenomenon that occurs when the mean free path of a gas particle is comparable to the characteristic length of the domain. Klinkenberg [40] first conducted the study of gas slippage in porous media, and found that due to gas slippage the permeability of gas $k_a$ (apparent permeability) through a tight porous medium is higher than that of liquid $k_d$ (intrinsic permeability). Beskok and Karniadakis [41] developed a second-order correlation between apparent permeability and intrinsic permeability

$$k_a = k_d f_c, \quad f_c = (1+\alpha(Kn)Kn)\left[1+\frac{4Kn}{1-bKn}\right] \tag{10}$$

where $f_c$ is correction factor and $b$ is slip coefficient which equals -1 for slip flow. $\alpha(Kn)$ is the rarefaction coefficient, which is calculated by [42]

$$\alpha(Kn) = \frac{1.358}{1+0.170Kn^{-0.4348}} \tag{11}$$

Eq. (10) has been shown to be capable of describing the four fluid flow regimes including viscous flow ($Kn<0.01$), slip flow($0.01<Kn<0.1$), transition flow($0.1< Kn <10$), and free molecular flow($Kn >10$). For calculating Kn, the mean free path $\lambda$ and the characteristic pore radius of the porous medium $r$ should be determined. The former one is calculated by [43]

$$\lambda = \frac{\mu}{p}\sqrt{\frac{\pi RT}{2M}} \tag{12}$$

where $R$ is the gas constant, $T$ the temperature and $M$ the molar mass. Following [44], the following expression proposed by Herd et al. is used to calculate $r$ [45]

$$r = 8.886\times10^{-2}\sqrt{\frac{k_d}{\varepsilon}} \tag{13}$$

in which the units of $r$ and $k$ are μm and mD ($1mD=9.869\times10^{-16}$ m$^2$), respectively.

*4.2 Pressure-dependent thickness of adsorption layer*

Due to the existence of the adsorption layer, the characteristic pore radius $r$ expressed by Eq. (13) will decrease. Based on the MD simulations in the present study and in Ref. [9], a pressure-dependent thickness of the adsorption layer

$$\delta = \frac{p-p_0}{p-p_1}(\delta_0 - \delta_1) \tag{14}$$

is subtracted from the pore radius. $\delta_0$ and $\delta_1$ are the thickness of the adsorbed layer under pressure $p_0$ and $p_1$, respectively. In the present study, $p_0 = 28$MPa, $\delta_0 = 0.7$nm, and $p_1 = 5$ MPa, $\delta_1 = 0$ nm.

*4.3 Simulation results and discussions*

Fluid flow between two parallel plates filled with a porous medium of porosity $\varepsilon$ is simulated. In the simulations, the domain is discretized by 200×80 square meshes. The relaxation time $\tau$ in LB is set as 0.9. No-slip boundary conditions are used for the top and bottom walls and a pressure difference is applied between left inlet and right outlet. For details of the simulation, one can refer to our previous study [19]. Fig. 12 shows the predicted intrinsic permeability as well as apparent permeability under different porosities and pressures. The intrinsic permeability is the permeability of liquid through the porous medium, in which slip is absent, and thus is not affected by the pressure. As expected, the intrinsic permeability decreases as the porosity of the porous medium decreases. The apparent permeability, which is normalized by the intrinsic permeability, however, presents different trends under different pressures. Under low pressure (1000psi), the normalized apparent permeability increases as the decrease of porosity, however, the trend of which is reverse for high pressure case (2000psi and 4000psi), and the higher the pressure, the smaller the normalized apparent permeability. This interesting result indicates that there are competitive factors affecting the normalized apparent permeability.

In nano/micro pores of shale matrix, gas transport through the pores with slippage on the solid surface. The slippage enhances as the *Kn* increases, and the stronger the slippage, the higher the apparent permeability. According to Eq. (12), a lower pressure leads to a longer mean free path, and thus a larger *Kn*, and a higher apparent permeability. On the other hand, adsorbed layer within the pores decreases the pore size, and thus reduces the permeability. Under a lower pressure, the adsorbed layer is thinner, and the decrease of the pore size is smaller. Therefore, both the effects of adsorbed layer and the slippage are affected by the pressure. Under a lower pressure, the enhancement of the slippage overwhelms the adverse effect of adsorbed layer on apparent permeability, and thus the normalized apparent increases as the porosity decreases. However, as the pressure increases, the favorable effect of slippage gradually becomes weaker, while the adverse effect of the adsorbed layer gets increasingly stronger. This leads to the decrease of the normalized apparent permeability with the decrease of the porosity. Such competitive effects were also found in a recent study of gas transport in a single pore of shale matrix [16]. Based on the LB results in Fig. 12, it can be concluded that adsorption effects should be included when studying gas flow in shale matrix.

**5 Conclusion**

Methane adsorption in nanosize pores of shale matrix and its effects on gas transport in shale matrix are investigated through multiscale simulations. Equilibrium MD simulations are performed to study the adsorption of methane on solid surface of nanopores in shale matrix within the typical values of pressure during shale gas extraction. Effects of the pore size, pressure and solid wall characteristics are explored in detail. The illumination of the MD simulation result on the LB simulations is discussed. The MD simulation results are upscaled into larger scale LB simulations where effects of the adsorption layer on the permeability of shale matrix is explored. The main conclusions of the present study are as follows

(1) Due to the adsorption effects of the solid wall, adsorbed layers are formed near the solid walls, the methane density in which is higher than that of the free gas region in the middle of the pores where pore pressure is defined. Under relatively low pressures, only one adsorbed layer is observed. As the pressure increases, a second adsorbed layer is gradually formed. The density ratio between the first/second adsorption layer and free gas decreases as the pressure increases, because the adsorption gradually approaches saturation as the pressure increases.

(2) When the width of a pore is less than 2.0 nm, methane molecules are always under the influence of adsorption from the pore wall, and adsorption with three density peaks can be observed. The densities in the adsorbed layer and the free gas region achieve constant values if the pore width is wider than a critical value under a certain pressure.

(3) Methane tends to adsorb on the organic matter surface. Besides, the crystal surface of calcite also affects the adsorption.

(4) Adsorbed layer affects the gas flow in pores of shale matrix through two ways: reducing the volume of void space and changing the slippage. MD simulation results reveal that a pressure-dependent thickness of the adsorbed layer should be adopted in larger scale LB simulations. MD simulations also find that adsorbed layer transports along the solid wall, in consistence with experiments in the literature. Effects of adsorbed layer on the slippage can be accounted for through the slip boundary conditions or SC pseudopotential model in LB simulations. Such effects, which are of great importance for the production of shale gas, need to be further studied.

(5) Mesoscopic LB simulations, which are based on a generalized Navier-Stokes equation for fluid flow through low-permeability porous media with slippage and adsorption effects, are conducted in a homogeneous shale matrix. It is found that both the effects of slippage and adsorption are pressure dependent. Under lower pressures, slippage is stronger and the volume of adsorption is less, leading to higher apparent permeability than intrinsic permeability. When the pressure is higher, the slippage becomes weaker while the volume reduction of void space due to adsorption gets larger, resulting in lower apparent permeability than intrinsic permeability.

**Acknowledgement**

The authors thank the support of National Nature Science Foundation of China (51406145 and 51136004). The authors also acknowledge the support of LANL's LDRD Program and Institutional Computing Program.

Petroleum Institute, New York, New York, USA, 1950, pp. 230-246.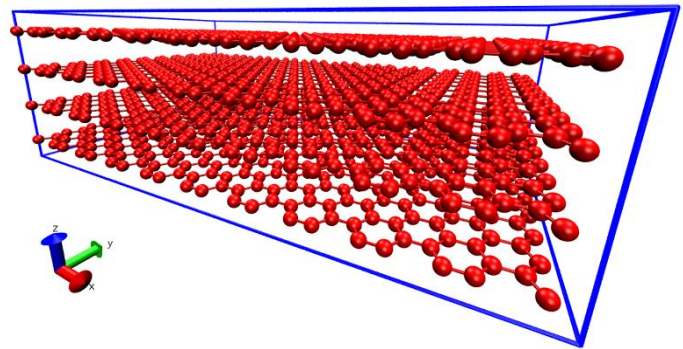

(a)

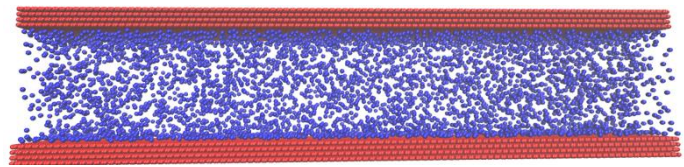

(b)

Fig. 1 Snapshot of the simulation box (a) the crystal structure of graphite (b) methane molecules in the graphite pore. Red represents carbon and blue represents methane.

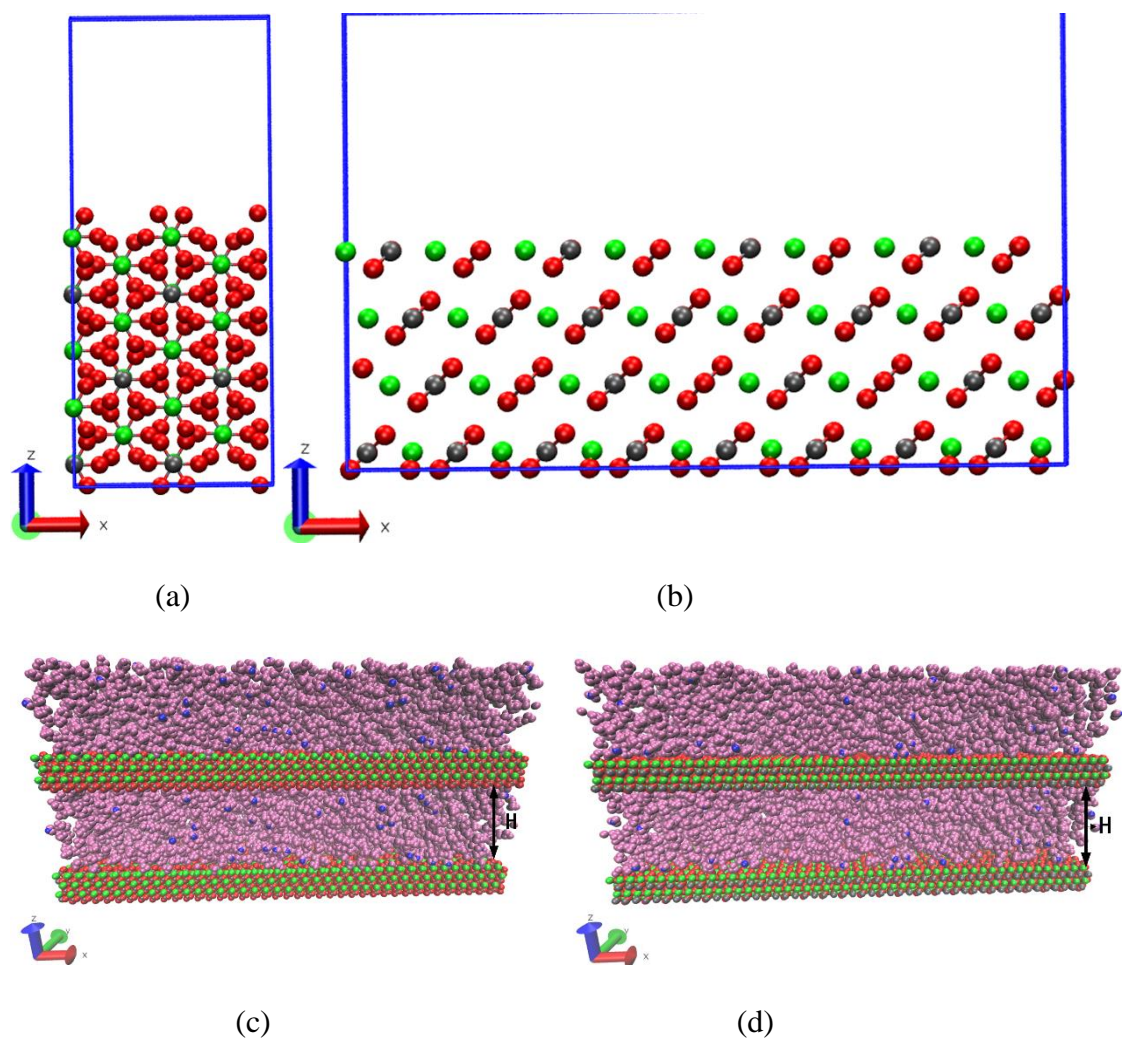

Fig. 2 Faces (1$\bar{1}$0) and (104) calcite crystal in (a) and (b), respectively (gray: C; red: O; green: Ca). Snapshot of the simulation box of methane (blue: C; ~~cyan~~mauve: H) adsorbed on calcite crystal surfaces (1$\bar{1}$0) and (104) in (c) and (d), respectively.

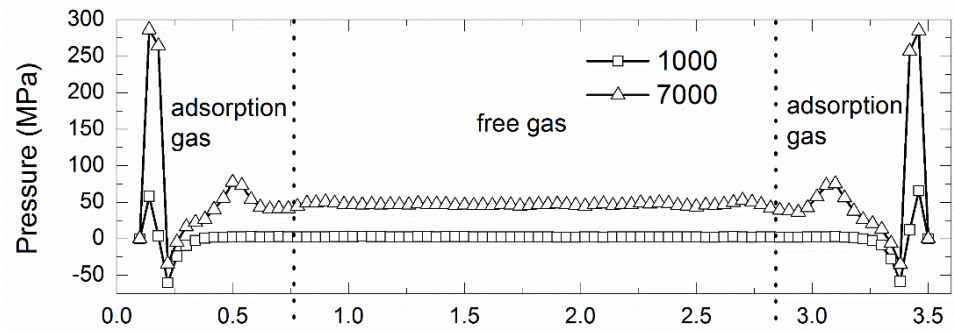

(a) Pressure

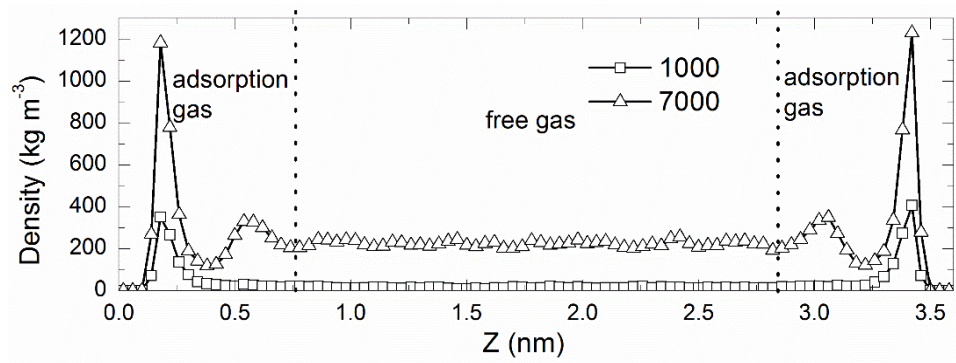

(b) Density

Fig 3 Pressure (a) and density (b) profile of methane in a graphite pore with pore width $H = 3.6$ nm.

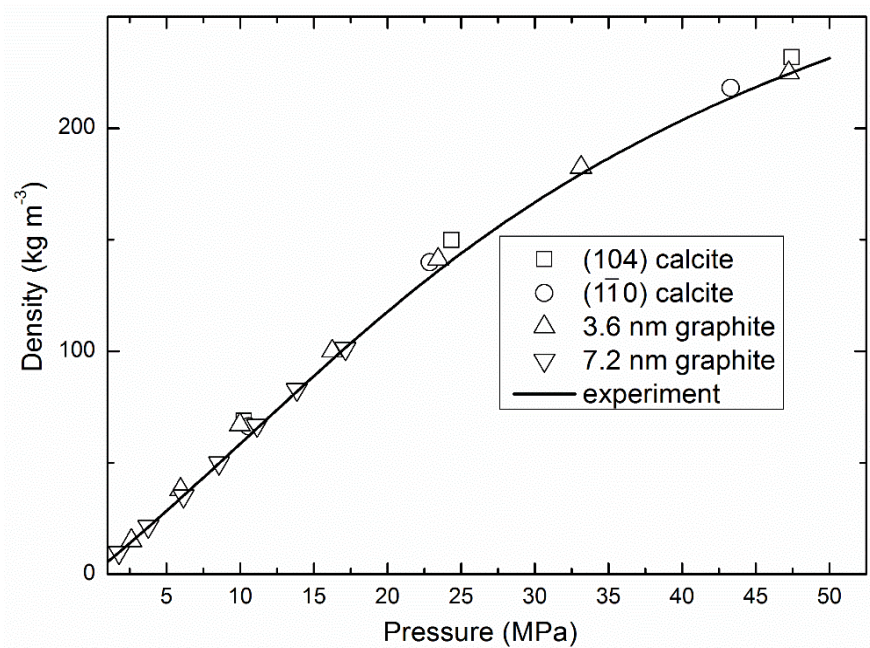

Fig 4 Comparison between MD simulations and experiment of density-pressure relationship for free gas

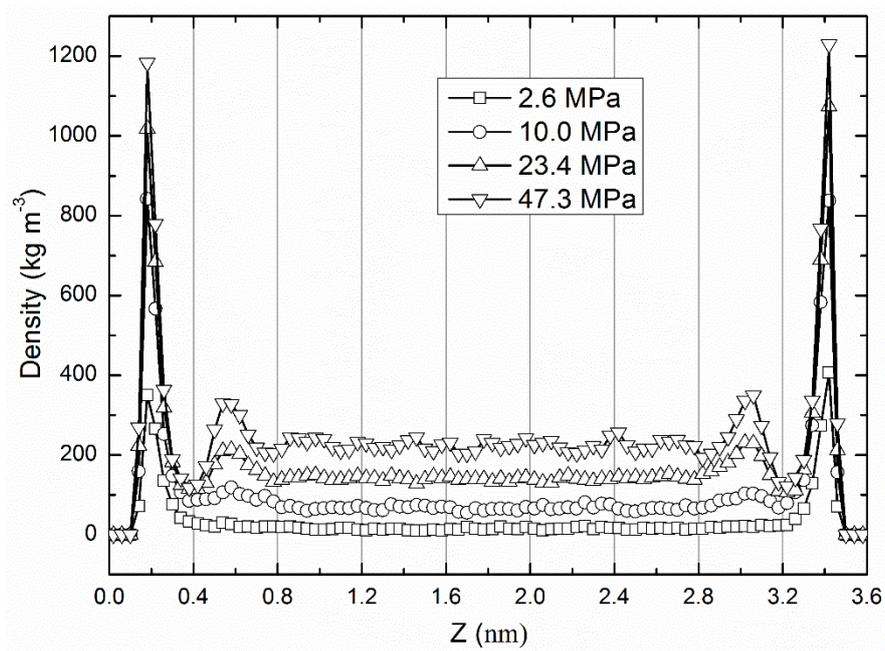

Fig 5 Density profile across the 3.6nm graphite pore under different pressures

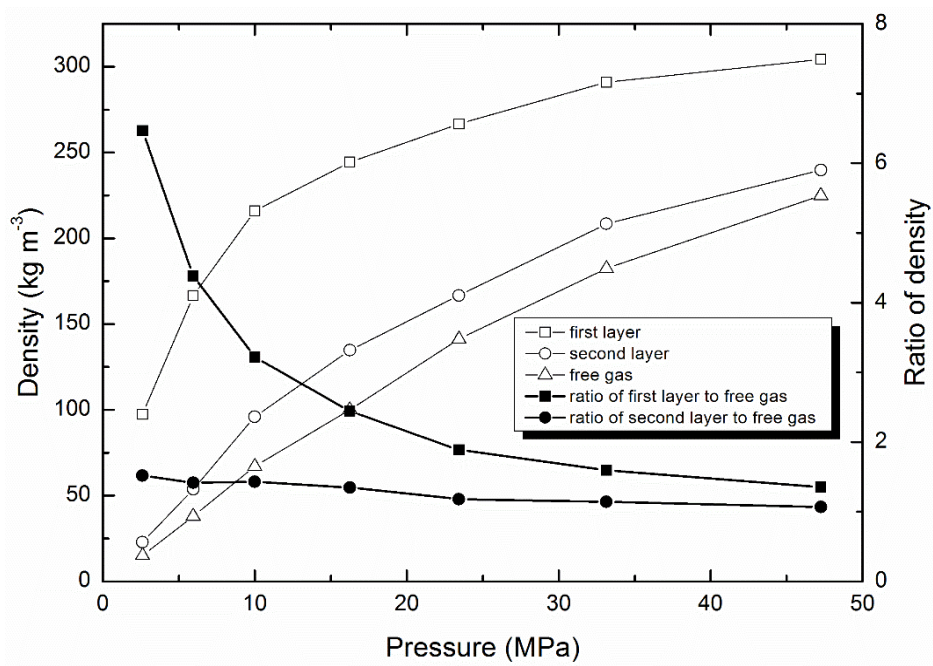

Figure 6 The average density of first layer, second layer and free gas in 3.6 nm graphite pore under different pressures. The open symbols represent the average density in different regions and the solid symbols represent the density ratios of different layers to free gas

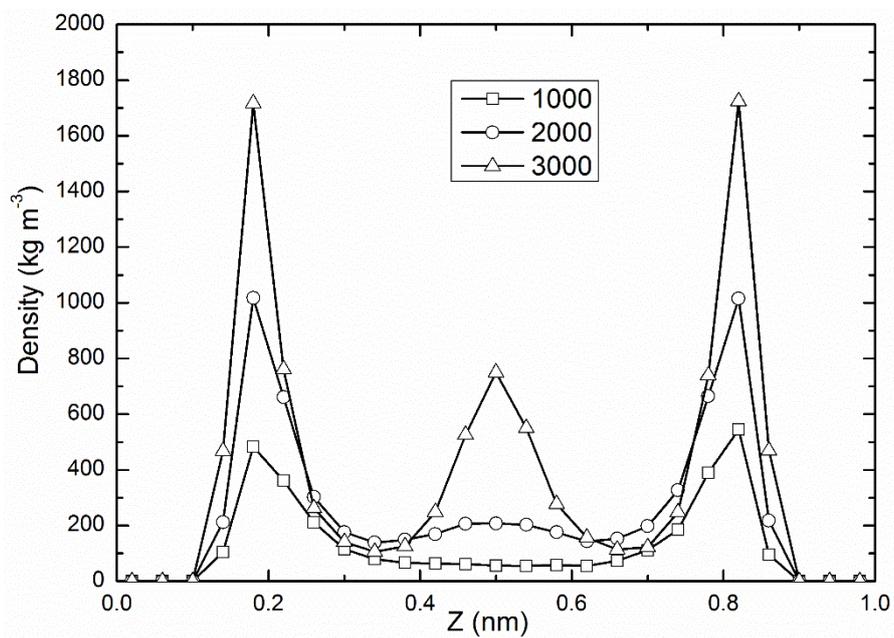

Fig 7 Density profile of methane in graphite pore with pore width H = 1.0 nm

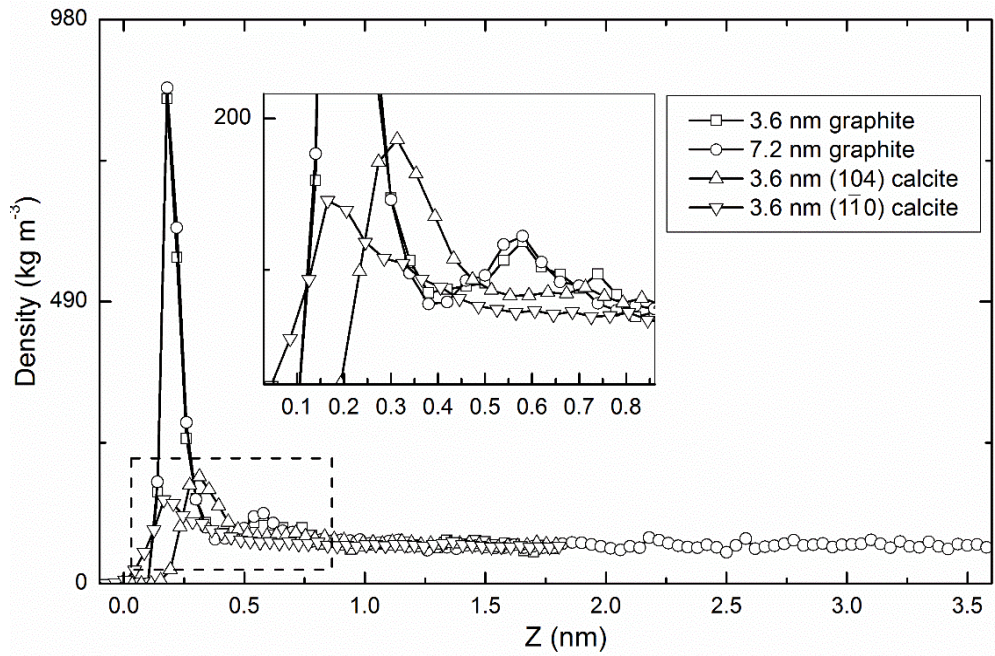

Fig 8 Density profile of methane across half-length of four different pore under similar pressure. The insert is the amplification of the zoom in dash square. □: 3.6 nm graphite pore with 9.9 MPa pressure; ○: 7.2 nm graphite pore with 11.15 MPa pressure; ∆: 3.6 nm (104) calcite pore with 10.25 MPa pressure; ▽: 3.6 nm ($1\bar{1}0$) calcite pore with 10.58 MPa pressure

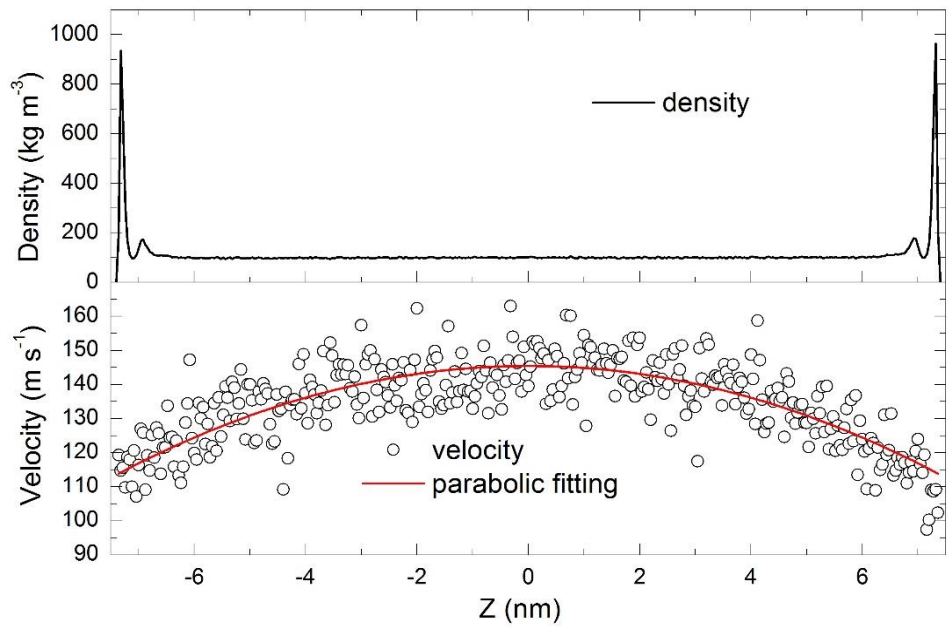

Figure 9 Density and velocity profile across the 15.0 nm pore of graphite.

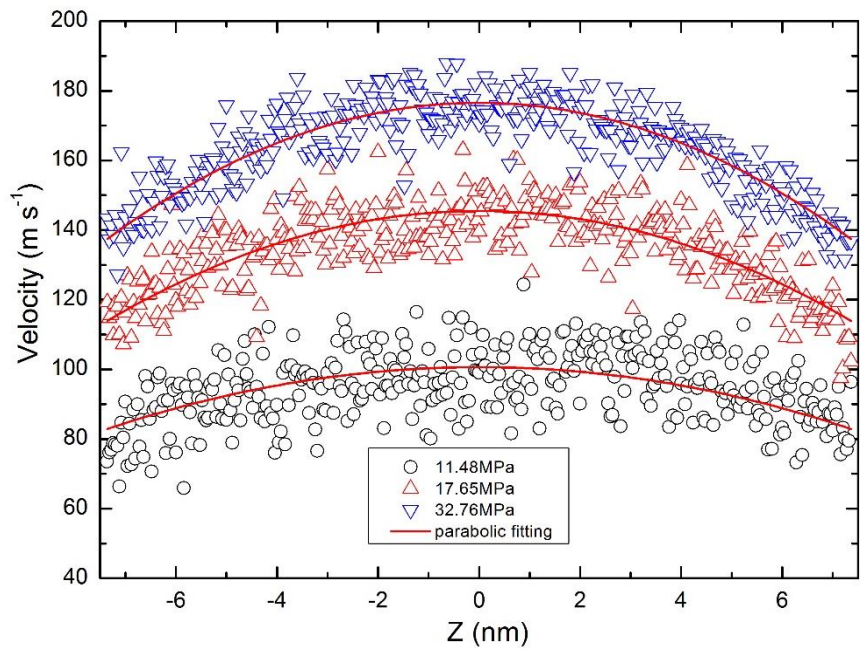

Figure 10 Velocity profile across the 15.0 nm pore at different pressures.

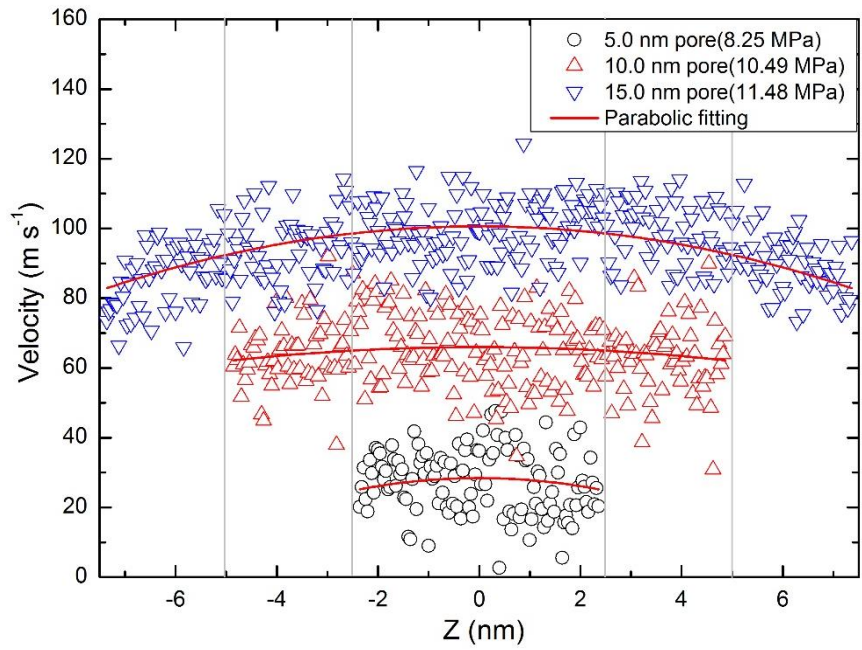

Figure 11 Velocity profile across the pores with different pore widths.

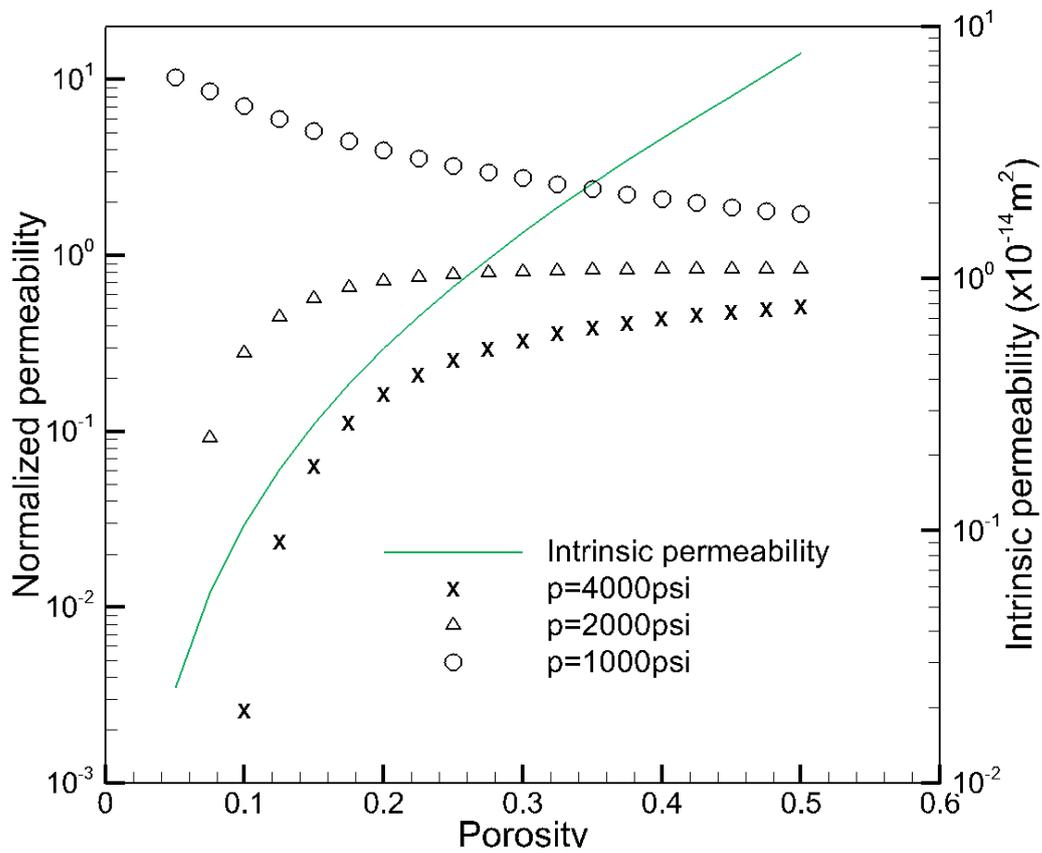

Fig. 12 Effects of adsorbed layer and slippage on the permeability under different pressures and porosities